\documentclass[prd,superscriptaddress,amsfonts,amssymb,amsmath,showpacs,twocolumn]{revtex4-2}
\usepackage{bm}
\usepackage{amsfonts}
\usepackage{latexsym}
\usepackage[latin1]{inputenc}
\usepackage{graphicx}
\usepackage{amsmath}
\usepackage{palatino}
\usepackage{mathpazo}
\usepackage{textcomp}
\usepackage{float}
\usepackage{booktabs}
\usepackage{dcolumn}
\usepackage{multirow}
\usepackage{hyperref}
\hypersetup{colorlinks,citecolor=blue}
\usepackage{amsmath}
\usepackage{xcolor}
\usepackage{orcidlink}
\usepackage[caption=false]{subfig}
\usepackage{commath}
\def\jnl@style{\it}
\def\aaref@jnl#1{{\jnl@style#1}}

\def\aaref@jnl#1{{\jnl@style#1}}

\def\aj{\aaref@jnl{AJ}}                   
\def\apj{\aaref@jnl{ApJ}}                 
\def\apjl{\aaref@jnl{ApJ}}                
\def\apjs{\aaref@jnl{ApJS}}               
\def\apss{\aaref@jnl{Ap\&SS}}             
\def\aap{\aaref@jnl{A\&A}}                
\def\aapr{\aaref@jnl{A\&A~Rev.}}          
\def\aaps{\aaref@jnl{A\&AS}}              
\def\mnras{\aaref@jnl{Mon.~Not.~Roy.~Astron.~Soc.}}             
\def\prd{\aaref@jnl{Phys.~Rev.~D}}        
\def\prc{\aaref@jnl{Phys.~Rev.~C}}  
\def\prl{\aaref@jnl{Phys.~Rev.~Lett.}}    
\def\qjras{\aaref@jnl{QJRAS}}             
\def\skytel{\aaref@jnl{S\&T}}             
\def\ssr{\aaref@jnl{Space~Sci.~Rev.}}     
\def\zap{\aaref@jnl{ZAp}}                 
\def\nat{\aaref@jnl{Nature}}              
\def\aplett{\aaref@jnl{Astrophys.~Lett.}} 
\def\apspr{\aaref@jnl{Astrophys.~Space~Phys.~Res.}} 
\def\physrep{\aaref@jnl{Phys.~Rep.}}      
\def\physscr{\aaref@jnl{Phys.~Scr}}       
\def\commat{\aaref@jnl{Comm.~Math.~Phys.}}              
\def\science{\aaref@jnl{Science}}               
\def\cqg{\aaref@jnl{Classical Quant.~Grav.}}            
\def\jpcs{\aaref@jnl{JPCS}}                                     
\def\ijmpd{\aaref@jnl{Int.~J.~Mod.~Phys.~D}}                    
\def\grg{\aaref@jnl{Gen.~Relat.~Gravit.}}               
\def\rpp{\aaref@jnl{Rep.~Prog.~Phys.}}          
\def\npa{\aaref@jnl{Nucl.~Phys.~A}}        
\def\lrr{\aaref@jnl{Living Rev.~Rel.}}                   
\def\jcap{\aaref@jnl{J.~Cosmology Astropart.~Phys.}}    
\def\rmp{\aaref@jnl{Rev.~Mod.~Phys.}}   
\def\epjc{\aaref@jnl{Eur.~Phys.~J.~C}}

\allowdisplaybreaks[1]

\begin{document}

\color{black}       

\title{Accelerating expansion of the universe in modified symmetric teleparallel gravity}

\author{Raja Solanki\orcidlink{0000-0001-8849-7688}}
\email{rajasolanki8268@gmail.com}
\affiliation{Department of Mathematics, Birla Institute of Technology and
Science-Pilani,\\ Hyderabad Campus, Hyderabad-500078, India.}
\author{Avik De\orcidlink{0000-0001-6475-3085}}
\email{de.math@gmail.com}
\affiliation{Department of Mathematical and Actuarial Sciences, Universiti Tunku Abdul Rahman, Jalan Sungai Long,
43000 Cheras, Malaysia}
\author{Sanjay Mandal\orcidlink{0000-0003-2570-2335}}
\email{sanjaymandal960@gmail.com}
\affiliation{Department of Mathematics, Birla Institute of Technology and
Science-Pilani,\\ Hyderabad Campus, Hyderabad-500078, India.}
\author{P.K. Sahoo\orcidlink{0000-0003-2130-8832}}
\email{pksahoo@hyderabad.bits-pilani.ac.in}
\affiliation{Department of Mathematics, Birla Institute of Technology and
Science-Pilani,\\ Hyderabad Campus, Hyderabad-500078, India.}

\date{\today}
\begin{abstract}

The fundamental nature and origin of dark energy are one of the premier mysteries of theoretical physics. In General Relativity Theory, the cosmological constant $\Lambda$ is the simplest explanation for dark energy. On the other hand, the cosmological constant $\Lambda$ suffers from a delicate issue so-called fine-tunning problem. This motivates one to modify the spacetime geometry of Einstein's GR. The $f(Q)$ gravity is a recently proposed modified theory of gravity in which the non-metricity scalar $Q$ drives the gravitational interaction. In this article, we consider a linear $f(Q)$ model, specifically $f(Q)=\alpha Q + \beta$, where $\alpha$ and $\beta$ are free parameters. Then we estimate the best fit values of model parameters that would be in agreement with the recent observational data sets. We use 31 points of the CC data sets, 6 points of the BAO data sets, and 1048 points from the Pantheon supernovae samples. We apply the Bayesian analysis and likelihood function along with the Markov Chain Monte Carlo (MCMC) method. Further, we analyze the physical behavior of cosmological parameters such as density, deceleration, and the EoS parameters corresponding to the constrained values of the model parameters. The evolution of deceleration parameter predicts a transition from decelerated to accelerated phases of the universe. Further, the evolution of equation of state parameter depicts quintessence type behavior of the dark energy fluid part. We find that our $f(Q)$ cosmological model can effectively describe the late time cosmic acceleration without invoking any dark energy component in the matter part. 

\end{abstract}

\maketitle

\section{Introduction}\label{sec1}

In the last two decades, evidences from Supernova searches \cite{Riess,Perlmutter}, WMAP experiment \cite{C.L.,D.N.}, CMBR  \cite{R.R.,Z.Y.}, LSS \cite{T.Koivisto,S.F.} and the BAO measurements \cite{D.J.,W.J.} indicate an accelerating cosmological expansion. The final fate of the universe is a topic of great concern. The root cause triggering this cosmological expansion is attributed to some sort of negative pressure dark energy (DE). The ultimate fate of our universe strongly depends on the fundamental nature of dark energy. The DE is usually characterized by an equation of state parameter $\omega_{DE} \equiv \frac{p_{DE}}{\rho_{DE}}$ that is a ratio of spatially homogeneous pressure $p_{DE}$ to the energy density $\rho_{DE}$ of dark energy. According to recent cosmological observations, the ambiguities are too large to differentiate among the cases : $\omega < -1$, $\omega = -1$, and $\omega > -1$. The value of the equation of state parameter for dark energy obtained by the WMAP  \cite{G.H.} which combined data from the $H_0$ measurements, supernovae, CMB, and BAO shows that $\omega=-1.084\pm0.063$ while in the year 2015, the Planck collaboration indicates that $\omega=-1.006\pm0.0451$ \cite{P15} and further in 2018 it reported that $\omega=-1.028\pm0.032$ \cite{N18}. The cosmological constant $\Lambda$ in GR is the simplest explanation for the dark energy and it is characterized by $\omega=-1$. However, there is a high discrepancy between the observed value of the cosmological constant $\Lambda$ and its expected value from quantum gravity \cite{S.W.}. This inconsistency in the value of $\Lambda$ is referred to as a cosmological constant problem. Another widely explored time varying DE model is the model with quintessence dark energy that is characterized by an equation of state $-1<\omega<-\frac{1}{3}$ \cite{RP,M.T.}. In such models, density of dark energy decreases with time \cite{LX}. Further, the least theoretically understood dark energy characterized by $\omega<-1$ is called phantom energy. The phantom energy case has gained much attention among theorists due to its strange properties. The phantom model represents growing dark energy that results in an extreme future expansion which leads to finite-time future singularity. For the classification of singularities, see the references \cite{KI,DB,DB-2}. It also violates all the four energy conditions that help to constrain wormholes \cite{ZH}. In this article, we follow a different mechanism in which dark energy evolves from the gravitational sector instead of the matter part. Such an approach has been widely used in the literature so-called modified theories of gravity \cite{L.A.,SA,R.F.}. Recently, $f(Q)$ gravity has been proposed by J. B. Jim\'enez et al \cite{J.B.} and it has gained much attention among cosmologists. The $f(R)$ theory of gravity is a generalization of GR in which the space-time is described by the non-vanishing curvature with vanishing torsion and non-metricity \cite{A.A.}. The $f(\mathcal{T})$ gravity theory is a generalization of the teleparallel equivalent of GR in which the space-time is described by the non-zero torsion with vanishing curvature and non-metricity \cite{GB}. Finally, the $f(Q)$ theory of gravity, which will be described in Section \ref{sec2}, is a generalization of the symmetric teleparallel equivalent of GR in which   the non-metricity scalar describes the gravitational interactions with zero curvature and torsion. Furthermore, the cosmological realisation of $f(R)$ theories remain close to those of GR, while models based on $f(T)$ theories suffer from strong coupling issues on generic FLRW backgrounds \cite{A.G}. This coupling problems observed in $f(T)$ theories are absent in $f(Q)$ models. Also, at the small scale quasi-static limit the predictions of the models based $f(T)$ and $f(Q)$ theory coincide whereas at higher scales model of $f(Q)$ theories generically propagate two scalar degrees of freedom that are absent in the case of $f(T)$. These two degrees of freedom disappear around maximally symmetric backgrounds and thus, cause the discussed strong coupling problem \cite{jimenez/2020}.

Although recently proposed, the $f(Q)$ gravity theory already presents some interesting and valuable applications in the literature. The first cosmological solutions in $f(Q)$ gravity appear in References \cite{jimenez/2020,khyllep/2021}, while $f(Q)$ cosmography and energy conditions can respectively be seen in \cite{mandal/2020,mandal/2020b}. The geodesic deviation equation in $f(Q)$ theory from its covariant formulation was studied \cite{Avv}. Quantum cosmology have been studied for a power-law model \cite{ND}. Cosmological solutions and growth index of matter perturbations have been investigated for a polynomial functional form of $f(Q)$ \cite{WK}. Isotropization process in an anisotropic Bianchi universe was investigated in a polynomial $f(Q)$ model \cite{iso}. Harko et al. analyzed the coupling matter in $f(Q)$ gravity by assuming a power-law function \cite{HRK}. The first evidence that non-metricity $f(Q)$ gravity can challenge $\Lambda$CDM has been presented in \cite{e1}. A set of constraints for $f(Q)$ cosmology with a $\Lambda$CDM background is presented in \cite{e2} using ground and space based gravitational waves observatories. Propagation of gravitational waves and redshift space distortions have been analyzed under the framework of $f(Q)$ gravity \cite{e3,e4}. A designer approach to $f(Q)$ gravity with signatures and its cosmological implications presented in \cite{e5,e6}. Furthermore, there are plenty of interesting works have been done in this theory such as black hole solution, wormhole solution, and general covariant symmetric teleparallel cosmology presented in \cite{e7,e8,e9}. Recently, some extensions of $f(Q)$ theory have been introduced in the literature namely $f(Q,T)$ theory \cite{m1} and Weyl $f(Q,T)$ theory \cite{m2}.

During last two decades the wealth of observational data increases. The majority of studies has been concentrated on evidences from Type Ia supernovae, baryon acoustic oscillations (BAO) and cosmic microwave background (CMB). The Hubble parameter dataset shows the intricate structure of the expansion of the universe. The ages of the most massive galaxies offer direct measurements of the Hubble parameter $H(z)$ at different redshifts $z$, resulting in the development of a new form of standard cosmological probe \cite{Jim2}. In the present manuscript we include 31 measurements of $H(z)$ spanned using differential age method which is treated as Cosmic chronometer (CC) data sets \cite{GS} in further study, BAO data consisting of six points \cite{BAO1}. Recently, Scolnic et al.  published a large Type Ia supernovae sample named Pantheon data sets consisting 1048 points that covers the redshift range $0.01< z< 2.3$ \cite{Scolnic/2018}. Our analysis uses the CC, BAO and Pantheon samples to constrain the cosmological model.

This manuscript is organized as follows. In Sec \ref{sec2}, we present the fundamental formulations in $F(Q)$ gravity. In Sec \ref{sec3}, Friedmann's equation for a flat cosmology along with a dark energy fluid part is discussed. In Sec \ref{sec4}, we consider a $F(Q)$ gravity model and derive the expressions for density, equation of state (EoS), and the deceleration parameter. In Sec \ref{sec5} we constraint the model parameters by using 31 points of the CC data sets, 6 points of the BAO data sets, and 1048 points from the Pantheon supernovae samples. Further, we investigate the physical behavior of cosmological parameters such as density, deceleration, and the EoS parameters corresponding to the constraint values of the model parameters. Finally, we discuss our conclusions in Sec \ref{sec6}. 

\section{Fundamental Formulations in $F(Q)$ Gravity Theory}\label{sec2}

The spacetime curvature demonstration of gravity is one of the most fundamental notions in natural sciences. Yet apart from the curvature, there are apparently two geometrical objects, namely, the torsion and the non-metricity of the spacetime's  geodesic structure which can establish the gravity. The standard theory of gravity governed by the General Relativity (GR) makes use of the spacetime curvature to determine gravity, with a torsion-free and metric-compatible connection, the very special Levi-Civita connection. As a byproduct, it assures that the geodesics are also autoparallel in this case. The second notion benefits from a metric-compatible, curvature-free connection with torsion, and is called the Teleparallel Equivalent of GR (TEGR). Whereas the last one determines gravity in a curvature and torsion-free environment, in which the non-metricity takes the complete charge of gravity, and is called the Symmetric Teleparallel Equivalent of GR (STEGR). As discussed in \cite{Tom}, the geometrical framework of this last setting is perhaps the simplest among the three equivalent theories of gravity because the connection can be globally completely cut off from the discussion by an appropriate choice of coordinates, called the coincident gauge coordinates. Moreover, in this particular coordinates system, it entails only the first derivatives of the metric tensor, unlike the standard theory of gravity, and thus generates a well-posed variational principle without any Gibbons-Hawking-York (GHY) boundary terms. 

A metric-affine spacetime equipped with a metric $g_{\mu\nu}$ that encodes angles and distances and a general affine connection $X^{\alpha}_{\mu\nu}$ that defines the notion of parallel transport and covariant derivatives. From differential geometry, it is well-known that the generic affine connection  $X^{\alpha}_{\mu\nu}$ admits a splitting into three parts \cite{Tom},

\begin{equation}\label{2a}
	X^\alpha_{\ \mu\nu}=\Gamma^\alpha_{\ \mu\nu}+K^\alpha_{\ \mu\nu}+L^\alpha_{\ \mu\nu},
\end{equation}

with the Levi-Civita connection of the metric tensor $g_{\mu\nu}$, 

\begin{equation}\label{2b}
	\Gamma^\alpha_{\ \mu\nu}\equiv\frac{1}{2}g^{\alpha\lambda}(g_{\mu\lambda,\nu}+g_{\lambda\nu,\mu}-g_{\mu\nu,\lambda})
\end{equation}

and the contortion tensor,

\begin{equation}\label{2c}
K^\alpha_{\ \mu\nu}\equiv\frac{1}{2}(T^{\alpha}_{\ \mu\nu}+T_{\mu \ \nu}^{\ \alpha}+T_{\nu \ \mu}^{\ \alpha})
\end{equation}
	
and the  distortion tensor,	
	
\begin{equation}\label{2d}
L^\alpha_{\ \mu\nu}\equiv\frac{1}{2}(Q^{\alpha}_{\ \mu\nu}-Q_{\mu \ \nu}^{\ \alpha}-Q_{\nu \ \mu}^{\ \alpha})
\end{equation}	

The last two terms are called torsion tensor and non-metricity tensor respectively and defined as 
 
\begin{equation}\label{2e}
 T^\alpha_{\ \mu\nu}\equiv X^\alpha_{\ \mu\nu}-X^\alpha_{\ \nu\mu}
\end{equation}
 and 
\begin{equation}\label{2f}
Q_{\alpha\mu\nu}\equiv\nabla_\alpha g_{\mu\nu}
\end{equation} 

The geometrical framework we use has a flat and torsionless connection so that it corresponds to pure coordinate transformation from the trivial connection as described in \cite{J.B.}. The connection can be parameterized as
 
\begin{equation}\label{2g}
X^\alpha \,_{\mu \beta} = \frac{\partial x^\alpha}{\partial \xi^\rho} \partial_ \mu \partial_\beta \xi^\rho.
\end{equation}
Here, $\xi^\alpha = \xi^\alpha (x^\mu)$ is an invertible relation. Hence, it is always possible to find a coordinate system so that the connection $ X^\alpha_{\ \mu\nu} $ vanishes. This situation is called coincident gauge and the covariant derivative $\nabla_{\alpha}$ reduces to the partial one $\partial_{\alpha} $. But in any other coordinate system in which this affine connection does not vanish, the metric evolution will be affected and result in a completely different theory \cite{AP,Avv}. Thus in the coincident gauge coordinate , we have

\begin{equation}\label{2h}
Q_{\alpha\mu\nu} = \partial_\alpha g_{\mu\nu}
\end{equation}
while in an arbitrary coordinate system,

\begin{equation}\label{2i}
Q_{\alpha\mu\nu}= \partial_\alpha g_{\mu\nu} - 2 X^\lambda_{\alpha (\mu} g_{\nu)\lambda}. 
\end{equation}

It is clear from the previous discussion that in STEGR under a coincident gauge coordinates, $X^{\alpha}_{\ \mu\nu}$ and $K^{\alpha}_{\ \mu\nu}$ vanish, and thus from equation (\ref{2a}) we can conclude that 
\begin{equation}\label{2j}
\Gamma^{\alpha}_{\ \mu\nu}=-L^{\alpha}_{\ \mu\nu}
\end{equation} 
using which we can calculate the required tensors and scalars in that specific coordinates.

The gravitational interactions in modified symmetric teleparallel geometry or $F(Q)$ gravity is governed by the following action:
\begin{equation}\label{2k}
S= \int{\frac{1}{2}F(Q)\sqrt{-g}d^4x} + \int{L_m\sqrt{-g}d^4x}
\end{equation}

Here $F(Q)$ is an arbitrary function of the non-metricity scalar $Q$,  $g=det( g_{\mu\nu} )$ and $L_m$ is the Lagrangian density of matter.

Due to the symmetry of the metric tensor $g_{\mu\nu}$, We can have only two independent traces from the non-metricity tensor $Q_{\alpha\mu\nu}$, namely,
 
\begin{equation}\label{2l}
Q_\alpha = Q_\alpha\:^\mu\:_\mu \: and\:  \tilde{Q}_\alpha = Q^\mu\:_{\alpha\mu}
\end{equation}

In addition, the non-metricity conjugate tensor is given by
\begin{equation}\label{2m}
4P^\lambda\:_{\mu\nu} = -Q^\lambda\:_{\mu\nu} + 2Q_{(\mu}\:^\lambda\:_{\nu)} + (Q^\lambda - \tilde{Q}^\lambda) g_{\mu\nu} - \delta^\lambda_{(\mu}Q_{\nu)}.
\end{equation}

The non-metricity scalar is acquired by \cite{LZ} 
\begin{equation}\label{2n}
Q = -Q_{\lambda\mu\nu}P^{\lambda\mu\nu}. 
\end{equation}

 Moreover, the stress-energy momentum tensor for the cosmic matter content is given by

\begin{equation}\label{2o}
\mathcal{T}_{\mu\nu} = \frac{-2}{\sqrt{-g}} \frac{\delta(\sqrt{-g}L_m)}{\delta g^{\mu\nu}}
\end{equation}

We denote $ F_Q = \frac{dF}{dQ} $ for the sake of convenience.

The field equation describing the gravitational interactions in $F(Q)$ gravity is obtained by varying the action \eqref{2k} with respect to the metric tensor as
\begin{widetext}
\begin{equation}\label{2p}
\frac{2}{\sqrt{-g}}\nabla_\lambda (\sqrt{-g}F_Q P^\lambda\:_{\mu\nu}) + \frac{1}{2}g_{\mu\nu}F+F_Q(P_{\mu\lambda\beta}Q_\nu\:^{\lambda\beta} - 2Q_{\lambda\beta\mu}P^{\lambda\beta}\:_\nu) = -T_{\mu\nu}.
\end{equation}
\end{widetext}

Moreover, one can obtain the following result by varying the action with respect to the connection,
\begin{equation}\label{2q}
\nabla_\mu \nabla_\nu (\sqrt{-g}F_Q P^{\mu\nu}\:_\lambda) =  0 
\end{equation}

\section{Flat FLRW Universe in $F(Q)$ Cosmology}\label{sec3}

Taking into account the homogeneity and isotropy of the universe, we describe our universe by the spatially flat FLRW line element \cite{Ryden} in Cartesian coordinates, which is, as a matter of fact also a coincident gauge coordinates, therefore from now connection becomes trivial and metric is the only fundamental variable,

\begin{equation}\label{3a}
ds^2= -dt^2 + a^2(t)[dx^2+dy^2+dz^2]
\end{equation}

Here, $ a(t) $ is a measure of the cosmic expansion at a cosmic time $t$, known as the scale factor . Now one can obtain the non-metricity scalar for the line element \eqref{3a} as

\begin{equation}\label{3b}
 Q= 6H^2  
\end{equation}

The stress-energy momentum tensor characterizing the matter-content of the universe by its matter-energy density $\rho$ and isotropic pressure $p$ for the line element \eqref{3a} is 

\begin{equation}\label{3c}
\mathcal{T}_{\mu\nu}=(\rho+p)u_\mu u_\nu + pg_{\mu\nu}
\end{equation}

Here $u^\mu=(1,0,0,0)$ are components of the four velocities of the perfect cosmic fluid.\\

Then the Friedmann like equations ruling the dynamics of the universe for the function $F(Q)=-Q+f(Q)$, are \cite{LZ}

\begin{equation}\label{3d}
f+Q-2Qf_Q = 2\rho
\end{equation}
and
\begin{equation}\label{3e}
\dot{H}=\frac{\rho+p}{2 \left(-1+f_Q+2Qf_{QQ} \right)} 
\end{equation}  

In particular, for the function $ F(Q)=-Q $, we can recover the usual Friedmann equations of GR.\\

Now, the trace of the field equations leads to the following matter conservation equation

\begin{equation}\label{3f}
\dot{\rho} + 3H\left(\rho+p\right)=0
\end{equation}

The equation of state that relates the usual pressure and matter-energy density of the cosmic fluid is given as \cite{Som}
\begin{equation}\label{3g}
p=\omega \rho
\end{equation}

Here $\omega$ is constant called equation of state (EoS) parameter.

The equation \eqref{3f} can be solved with the help of equation \eqref{3g}, to give 

\begin{equation}\label{3h}
\rho=\rho_0 a(t)^{-3\left(1+\omega \right)}
\end{equation}

These equations \eqref{3d} and \eqref{3e} can be interpreted as symmetric teleparallel equivalent to GR (STG) cosmology with an additional component coming due to non-metricity of space-time that behaves like dark energy fluid part. These dark energy components coming due to non-metricity are defined by 

\begin{equation}\label{3i}
\rho_{DE}=-\frac{f}{2}+Qf_Q
\end{equation}

and 

\begin{equation}\label{3j}
p_{DE}= -\rho_{DE} - 2\dot{H} \left(f_Q+2Qf_{QQ} \right)
\end{equation}

Now, the dimensionless density parameter for the dark energy fluid part is defined as

\begin{equation}\label{3k}
\Omega_{DE}= \frac{\rho_{DE}}{3H^2}
\end{equation}

Moreover, the equation of state (EoS) parameter that relates the energy density and pressure of the dark energy component is

\begin{equation}\label{3l}
\omega_{DE}= \frac{p_{DE}}{\rho_{DE}} = -1 + 4\dot{H} \left( \frac{f_Q+2Qf_{QQ}}{f-2Qf_Q} \right)
\end{equation}

Using \eqref{3d} and \eqref{3e}, we have

\begin{equation}\label{3m}
\omega_{DE} = -1 + \left(1+\omega\right) \frac{\left(f+Q-2Qf_Q\right) \left(f_Q+2Qf_{QQ} \right)}{\left(-1+f_Q+2Qf_{QQ} \right) \left(f-2Qf_Q \right)}
\end{equation}

Thus, the effective Friedmann equations along with a dark energy fluid part coming due to non-metricity reads

\begin{equation}\label{3n}
H^2=\frac{1}{3} \left[ \rho+\rho_{DE} \right]
\end{equation}

\begin{equation}\label{3o}
\dot{H}=-\frac{1}{2} \left[ \rho+p+\rho_{DE}+p_{DE} \right]
\end{equation}

Furthermore, these dark energy components satisfy the standard continuity equation

\begin{equation}\label{3p}
\dot{\rho}_{DE} + 3H \left( \rho_{DE} + p_{DE} \right) = 0
\end{equation}

\section{Cosmological $F(Q)$ Model}\label{sec4}

Despite aforementioned issues, cosmological constant $\Lambda$ in GR is most successful model so far, therefore this motivate us to consider following linear $f(Q)$ model \cite{GM}, 
 
\begin{equation}\label{4a}
f(Q)= \alpha Q + \beta
\end{equation}
 
Here $\alpha$ and $\beta$ are free model parameters.\\

Then for this particular $f(Q)$ model, we have a first-order differential equation for a universe consisting of non-relativistic pressureless matter reads as

\begin{widetext}
\begin{equation}\label{4b}
\dot{H} \left[ \alpha-1 \right] + \frac{3}{2} H^2  \left[ \alpha-1+ \frac{\beta}{6H^2} \right]=0
\end{equation}
\end{widetext}

By using equations \eqref{3i} and \eqref{4a}, we obtain

\begin{equation}\label{4c}
\rho_{DE}= 3\alpha H^2 - \frac{\beta}{2}
\end{equation}

Again, by using equations \eqref{3k} and \eqref{3m}, we obtain

\begin{equation}\label{4d}
\Omega_{DE}= \frac{\rho_{DE}}{3H^2} = \alpha - \frac{\beta}{6H^2}
\end{equation}
 
and

\begin{equation}\label{4e}
\omega_{DE}= -1+ \frac{\left( \alpha-1- \frac{\beta}{6H^2} \right) \alpha}{\left( \alpha-1 \right)\left( \alpha - \frac{\beta}{6H^2} \right)} 
\end{equation}

Thus, the effective EoS parameter for our model is 

\begin{equation}\label{4f}
w_{eff}= \frac{p_{eff}}{\rho_{eff}}=\frac{p_{DE}}{\rho + \rho_{DE}}
\end{equation}

where, $p_{eff}$ and $\rho_{eff}$ correspond to the total pressure and energy density of the universe. Then we have

\begin{widetext}
\begin{equation}\label{4g}
w_{eff} = - \left( \alpha - \frac{\beta}{6H^2}  \right) +  \frac{\left( \alpha-1- \frac{\beta}{6H^2} \right) \alpha }{\left( \alpha-1 \right)} 
\end{equation}
\end{widetext}

The deceleration parameter is a key component to describe the expansion phase of the universe, it is defined as \cite{Muj}

\begin{equation}\label{4h}
q= \frac{1}{2} \left( 1+3\Omega_{DE} \omega_{DE} \right)
\end{equation}

Using \eqref{4d} and \eqref{4e}, we get

\begin{widetext}
\begin{equation}\label{4i}
q=\frac{1}{2} + \frac{3}{2} \bigl\{ - \left( \alpha - \frac{\beta}{6H^2}  \right) +  \frac{\left( \alpha-1- \frac{\beta}{6H^2} \right) \alpha }{\left( \alpha-1 \right)}   \bigr\}
\end{equation}
\end{widetext}

Now by solving equation \eqref{4b}, we obtained expression for Hubble parameter in terms of redshift as

\begin{equation}\label{4j}
H(z)= \bigl\{ H_0^2(1+z)^3+ \frac{\beta}{6(\alpha-1)} \left[ 1-(1+z)^3 \right] \bigr\}^{\frac{1}{2}}
\end{equation}

where $H_0=67.9\pm 0.5$ km/s/Mpc \cite{planck_collaboration/2020} is the present value of the Hubble parameter. Now, our aim is to constraint the free parameters using observational data.

\section{Observational Constraints}\label{sec5}

To constrain the model parameters of our cosmological model, we use the most recent Cosmic chronometer data set, BAO data set, and Supernovae observations. We use 31 points of the CC data sets, 6 points of the BAO data sets, and 1048 points from the Pantheon supernovae samples. We apply the Bayesian analysis and likelihood function along with the Markov Chain Monte Carlo (MCMC) method in \texttt{emcee} python library \cite{Mackey/2013}. 

\subsubsection{Cosmic Chronometer datasets}

The Hubble parameter can be expressed as $H(z)=-dz/[dt(1+z)]$. As dz is derived from a spectroscopic survey, the model-independent value of the Hubble parameter may be calculated by measuring dt. Here we use the collection of $H(z)$ data points that consists of 31 points measured from differential age technique (listed in \cite{cc}). To calculate mean values of the model parameters $\alpha$ and $\beta$, we have taken the chi-square function as:

\begin{equation}
\chi _{H}^{2}(\alpha,\beta)=\sum\limits_{k=1}^{31}
\frac{[H_{th}(z_{k},\alpha,\beta)-H_{obs}(z_{k})]^{2}}{
\sigma _{H(z_{k})}^{2}}.  \label{5a}
\end{equation}

Here, $H_{th}$ represents theoretical value of the Hubble parameter predicted by our cosmological model while $H_{obs}$ represents its observed value and $\sigma_{H(z_{k})}$ is the standard error in the observed value of $H$.

%

\subsubsection{Pantheon datasets}

Initially, the observational studies on sample of 50 points of type Ia supernovae revealed that our universe is accelerating. In the last two decades, studies on more sample of type Ia supernovae data sets has been increased. In this manuscript, we have used a sample of 1048 spectroscopically confirmed type Ia supernovae known as Pantheon data sets. Scolnic et al. \cite{Scolnic/2018} put together the Pantheon samples consisting of 1048 type Ia supernovae in the redshift range $0.01 < z < 2.3$. The PanSTARSS1 Medium Deep Survey, SDSS, SNLS and numerous low-z, and HST samples contribute to it. For a flat universe \cite{planck_collaboration/2020}, the luminosity distance is given by 

\begin{equation}\label{5b}
D_{L}(z)= (1+z) \int_{0}^{z} \frac{c dz'}{H(z')},
\end{equation}
Here $c$ is the speed of light.

The $\chi^{2}$ function for type Ia supernovae is obtained by correlating the theoretical distance modulus 

\begin{equation}\label{5c}
\mu(z)= 5log_{10}D_{L}(z)+\mu_{0}, 
\end{equation}
with 
\begin{equation}\label{5d}
\mu_{0} =  5log(1/H_{0}Mpc) + 25,
\end{equation}
such that
\begin{equation}\label{5e}
\chi^2_{SN}(p_1,....)=\sum_{i,j=1}^{1048}\bigtriangledown\mu_{i}\left(C^{-1}_{SN}\right)_{ij}\bigtriangledown\mu_{j},
\end{equation}
where $p_j$ represents the assumed model's free parameters and $C_{SN}$ is the covariance metric \cite{Scolnic/2018}, and
 \begin{align*}
  \quad \bigtriangledown\mu_{i}=\mu^{th}(z_i,p_1,...)-\mu_i^{obs}.
 \end{align*}
Here $\mu_{th}$ represents theoretical value of the distance modulus, $\mu_{obs}$ represents its observed.

%

\subsubsection{BAO datasets}

The Baryonic Acoustic Oscillation (BAO) data set consists of 6dFGS, SDSS and WiggleZ surveys that comprise BAO measurements at six different redshifts in table-1. The characteristic scale of BAO is governed by the sound horizon $r_{s}$ at the photon decoupling epoch $z_{\ast }$ using the following relation, 

\begin{equation}\label{5f}
r_{s}(z_{\ast })=\frac{c}{\sqrt{3}}\int_{0}^{\frac{1}{1+z_{\ast }}}\frac{da}{
a^{2}H(a)\sqrt{1+(3\Omega _{0b}/4\Omega _{0\gamma })a}}
\end{equation}

Here $\Omega _{0b}$ and $\Omega _{0\gamma }$ represent present densities of baryons and photons respectively.

In this work, BAO datasets of six points for $d_{A}(z_{\ast })/D_{V}(z_{BAO})$ is taken from the references \cite{BAO1, BAO2, BAO3, BAO4, BAO5, BAO6}. Here the redshift at the photon decoupling epoch is taken as $z_{\ast }\approx 1091$ and  $d_{A}(z)=\int_{0}^{z}\frac{dz^{\prime }}{H(z^{\prime })} $ is the co-moving angular diameter
distance together with the dilation scale $D_{V}(z)=\left(
d_{A}(z)^{2}z/H(z)\right) ^{1/3}$. The chi square function for the BAO distance datasets is used as \cite{BAO6}
 
\begin{equation}\label{5g}
\chi _{BAO}^{2}=X^{T}C^{-1}X\,,  
\end{equation}

\begin{widetext}
\begin{table}[H]
\begin{center}
\begin{tabular}{|c|c|c|c|c|c|c|}
\hline
$z_{BAO}$ & $0.106$ & $0.2$ & $0.35$ & $0.44$ & $0.6$ & $0.73$ \\ \hline
$\frac{d_{A}(z_{\ast })}{D_{V}(z_{BAO})}$ & $30.95\pm 1.46$ & $17.55\pm 0.60$
& $10.11\pm 0.37$ & $8.44\pm 0.67$ & $6.69\pm 0.33$ & $5.45\pm 0.31$ \\ 
\hline
\end{tabular}
\caption{Values of $d_{A}(z_{\ast })/D_{V}(z_{BAO})$ for distinct values of $z_{BAO}$}
\end{center}
\label{Table-1}
\end{table}
\end{widetext}

where 

\begin{equation*}
X=\left( 
\begin{array}{c}
\frac{d_{A}(z_{\star })}{D_{V}(0.106)}-30.95 \\ 
\frac{d_{A}(z_{\star })}{D_{V}(0.2)}-17.55 \\ 
\frac{d_{A}(z_{\star })}{D_{V}(0.35)}-10.11 \\ 
\frac{d_{A}(z_{\star })}{D_{V}(0.44)}-8.44 \\ 
\frac{d_{A}(z_{\star })}{D_{V}(0.6)}-6.69 \\ 
\frac{d_{A}(z_{\star })}{D_{V}(0.73)}-5.45%
\end{array}%
\right) \,,
\end{equation*}

and $C^{-1}$ is the inverse covariance matrix reads as 
\cite{BAO6}. 

\begin{widetext}
\begin{equation*}
C^{-1}=\left( 
\begin{array}{cccccc}
0.48435 & -0.101383 & -0.164945 & -0.0305703 & -0.097874 & -0.106738 \\ 
-0.101383 & 3.2882 & -2.45497 & -0.0787898 & -0.252254 & -0.2751 \\ 
-0.164945 & -2.454987 & 9.55916 & -0.128187 & -0.410404 & -0.447574 \\ 
-0.0305703 & -0.0787898 & -0.128187 & 2.78728 & -2.75632 & 1.16437 \\ 
-0.097874 & -0.252254 & -0.410404 & -2.75632 & 14.9245 & -7.32441 \\ 
-0.106738 & -0.2751 & -0.447574 & 1.16437 & -7.32441 & 14.5022%
\end{array}%
\right) \,.
\end{equation*}
\end{widetext}

\begin{figure}
\includegraphics[scale=0.85]{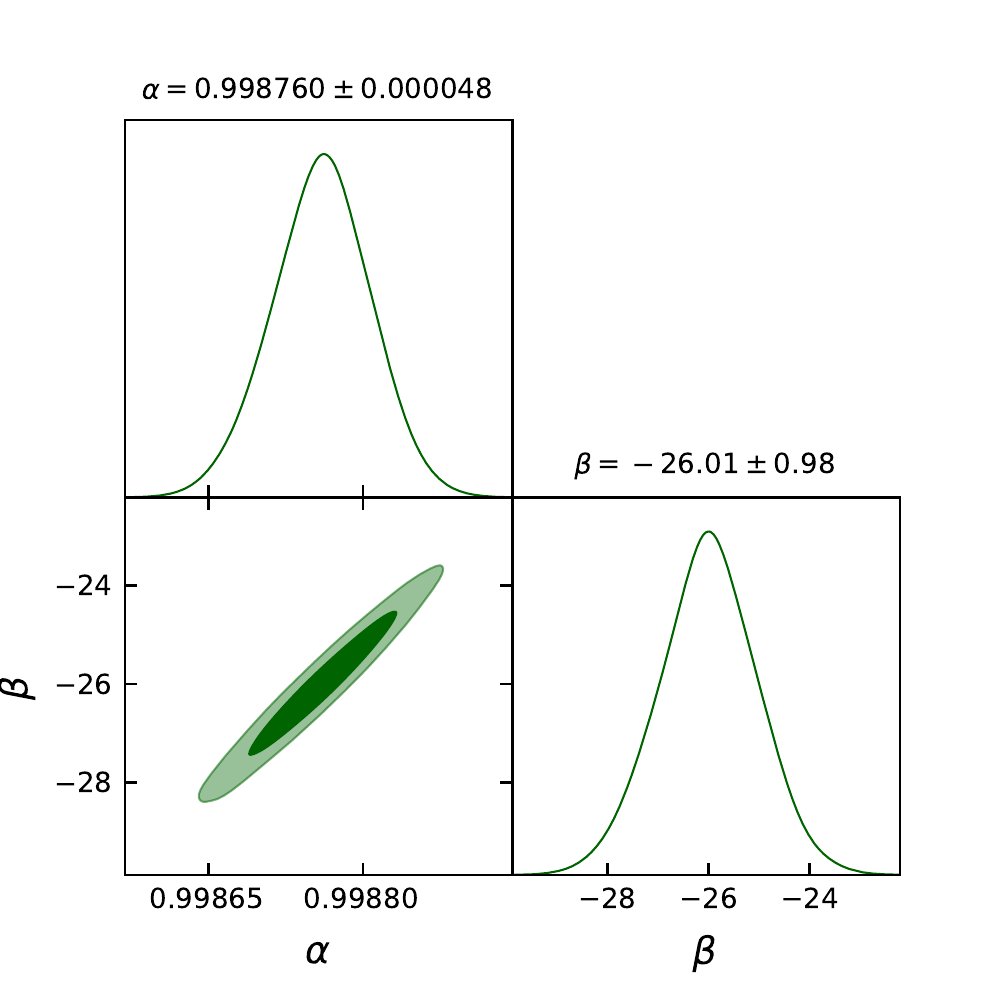}
\caption{The $1-\sigma$ and $2-\sigma$ likelihood contours for the model parameters using the combination CC+BAO+Pantheon data sets.}\label{f3}
\end{figure}

We have calculated the best fit ranges for parameters $\alpha$ and $\beta$ of our cosmological model by minimizing the chi-square function for the combination $CC+BAO+Pantheon$. The obtained best fit values of the model parameters are $\alpha= 0.998760 \pm 0.000048$ and $\beta=-26.01 \pm 0.98$. The $1-\sigma$ and $2-\sigma$ likelihood contour for the model parameters using the combine CC+BAO+Pantheon data sets is presented in Fig \ref{f3}.

\subsubsection{Cosmological Parameters}

The evolution of different cosmological parameters of our cosmological model such as density, deceleration, and the EoS parameters corresponding to the constrained values of the model parameters are presented below.

\begin{figure}[H]
\includegraphics[scale=0.6]{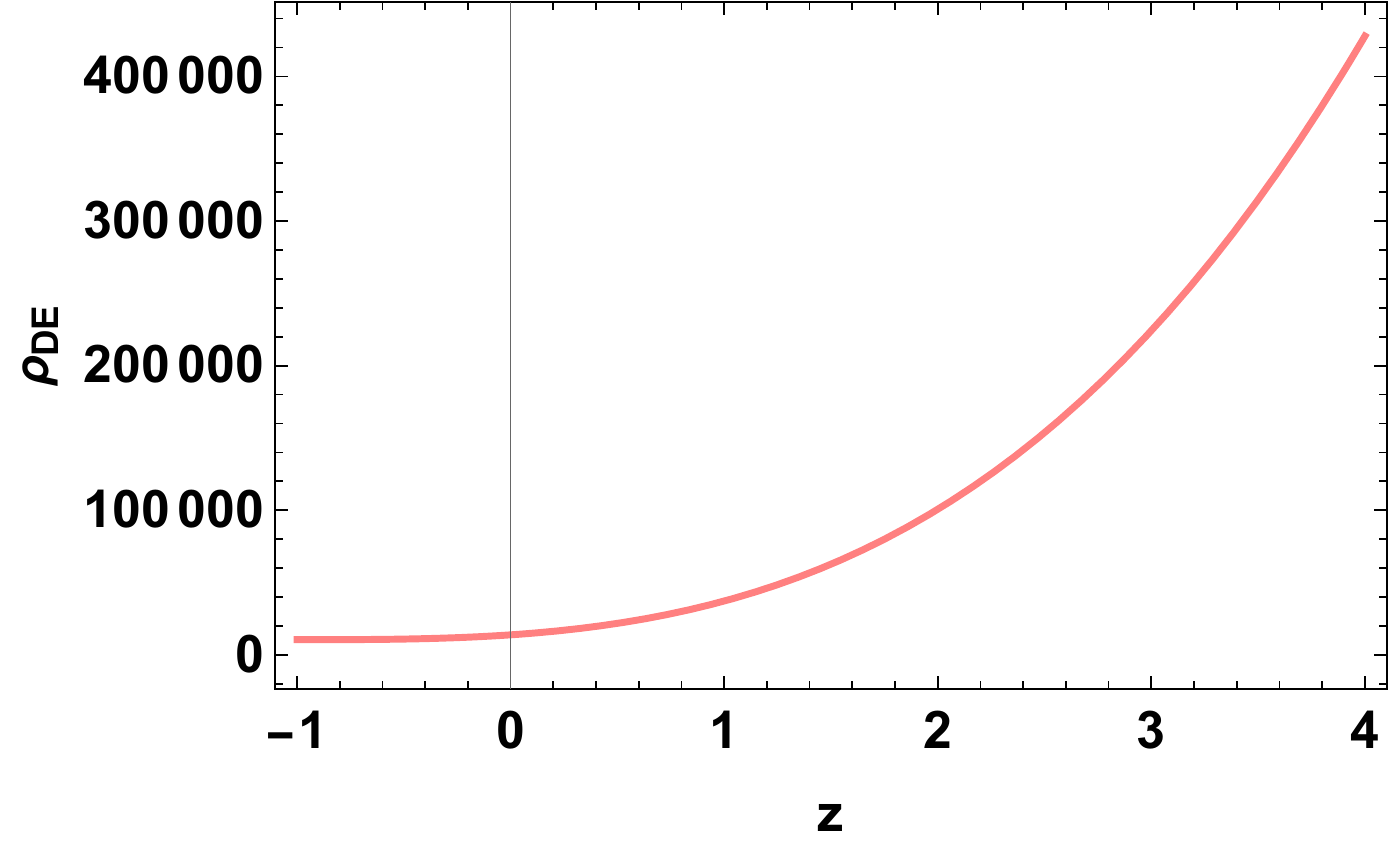}
\caption{Profile of the density of the dark energy component vs redshift z .}\label{f4}
\end{figure}


\begin{figure}[H]
\includegraphics[scale=0.57]{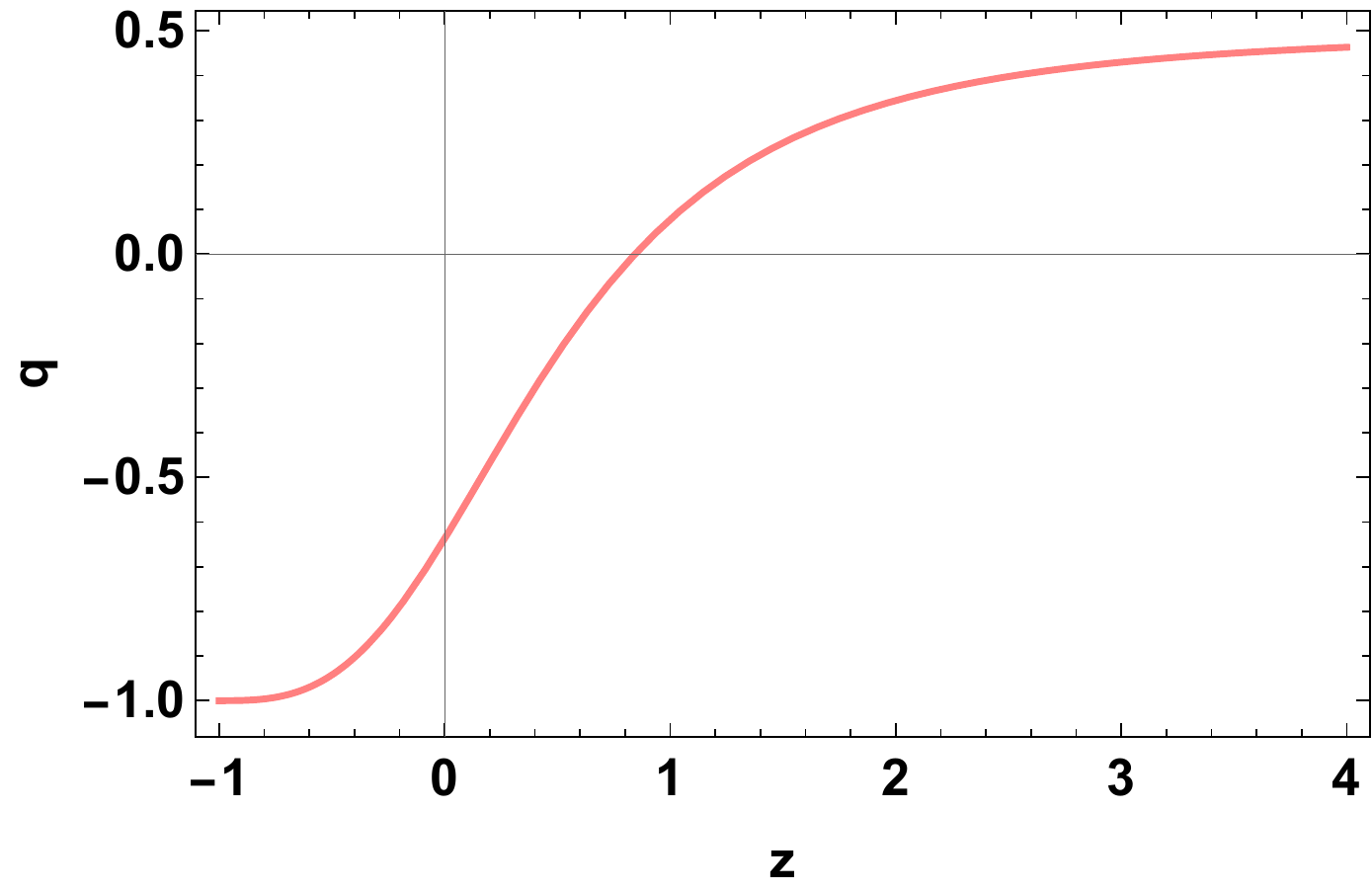}
\caption{Profile of the deceleration parameter vs redshift z .}\label{f5}
\end{figure}

\begin{figure}[H]
\includegraphics[scale=0.57]{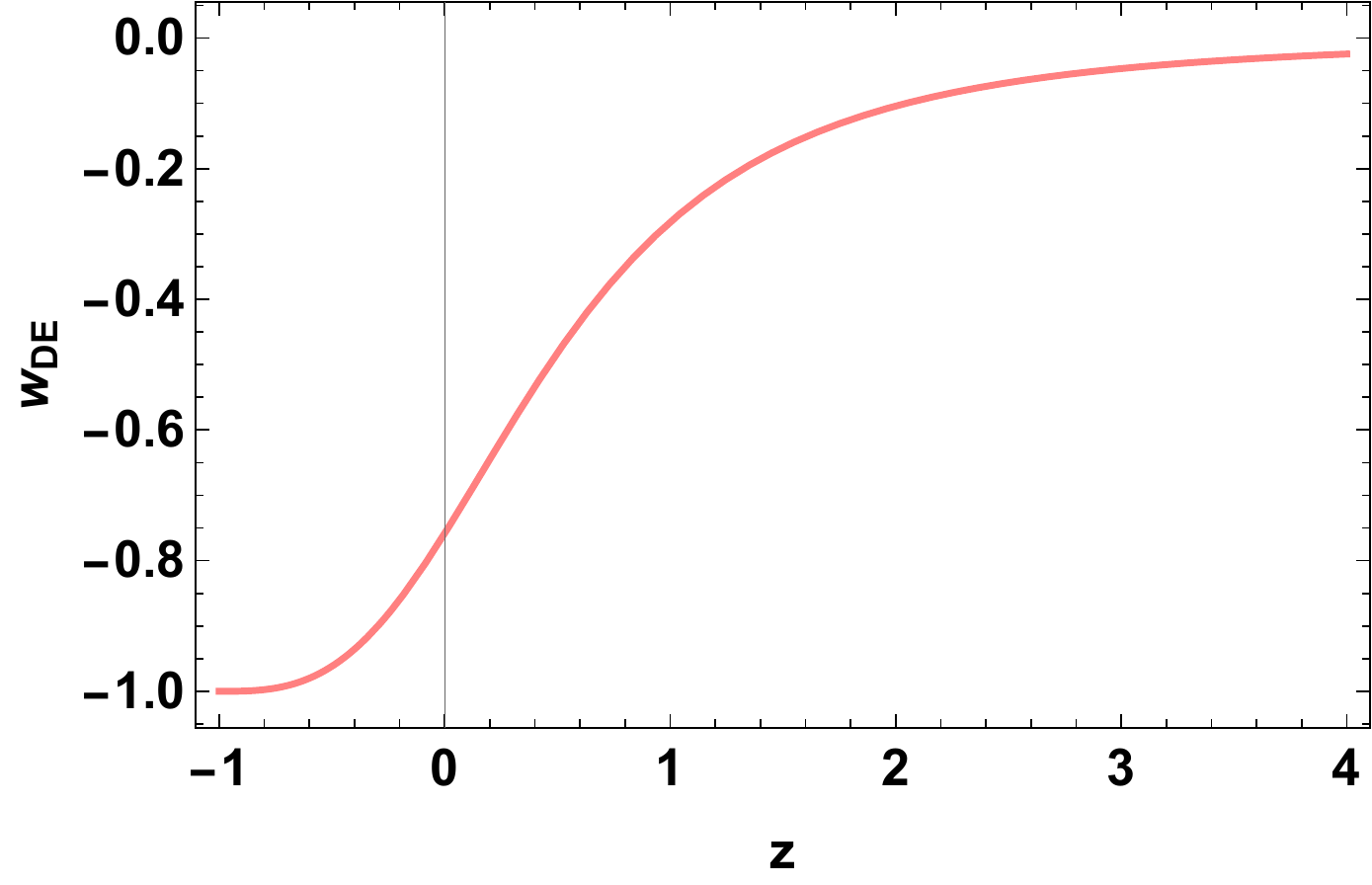}
\caption{Profile of the EoS parameter for the dark energy component vs redshift z .}\label{f6}
\end{figure}

\begin{figure}[H]
\includegraphics[scale=0.57]{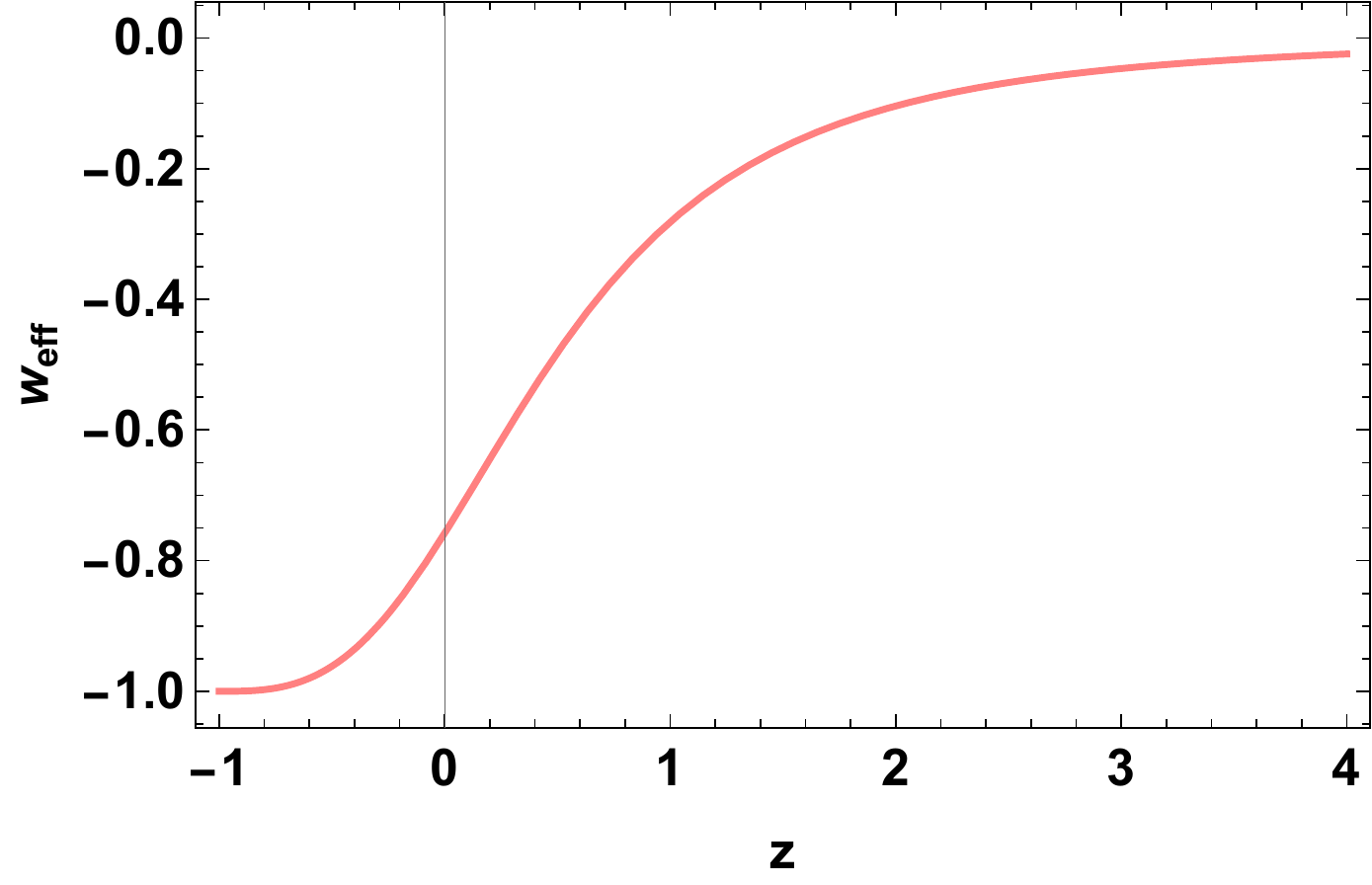}
\caption{Profile of the effective EoS parameter vs redshift z .}\label{f7}
\end{figure}

Form fig. \ref{f4} it is clear that the energy density of the dark energy component of the universe decreases with cosmic time and it falls off to be zero in the far future. Hence our model represents a decaying dark energy model. Fig. \ref{f5} show that a transition from decelerating to accelerating phase has been experienced by the universe in the recent past with transition redshift $z_t=0.844^{+0.0027}_{-0.0025}$. Further from fig. \ref{f6} snd \ref{f7}, we observe that the EoS parameter for the dark energy fluid evolving due to non-metricity and effective EoS parameter shows quintessence like behavior. The present values of the deceleration parameter and EoS parameter for the dark energy fluid part correspond to the constrained values of the model parameters are $q_0=-0.63^{+0.0011}_{-0.0012}$ and $w_0=-0.75^{+0.0007}_{-0.0008}$.

\section{Conclusion}\label{sec6}

Understanding the evolution of dark energy is a great challenge for modern cosmology. Dark energy occupies nearly 68.3\% of the entire universe while dark matter and baryonic matter occupy nearly 26.8\% and 4.9\% of the total energy content of the universe. However, the vacuum quantum energy can well describe the origin of dark energy by means of the cosmological constant $\Lambda$ in the field equations of GR. Although the aforementioned issues
related to $\Lambda$ supply the search for alternative explanations for the dark energy candidate.

In this article, we attempted to describe the evolution of dark energy from the geometry of spacetime. We considered a linear $f(Q)$ model $f(Q)=\alpha Q + \beta $, where $\alpha$ and $\beta$ are free parameters. Then we found the expressions for density, deceleration, and the EoS parameters for our cosmological model. Further, to constrain the model parameters we used CC data sets consisting 31 data points, 6 points of the BAO data sets, and 1048 points from the Pantheon supernovae samples. We have calculated the best fit ranges of the model parameters $\alpha$ and $\beta$ for the combine $CC+BAO+Pantheon$ data sets. The obtained best fit values of the model parameters are $\alpha= 0.998760 \pm 0.000048$ and $\beta=-26.01 \pm 0.98$.  Further, we have studied the evolution of different cosmological parameters corresponding to these best fit values of the model parameters. The evolution trajectory of the deceleration parameter shows that our universe had experienced a transition from deceleration to acceleration phase in the recent past with the transition redshift $z_t=0.844^{+0.0027}_{-0.0025}$. The obtained present value of the deceleration parameter is $q_0=-0.63^{+0.0011}_{-0.0012}$. However, in case of standard $\Lambda$CDM model, the present day value of the deceleration parameter is $q_0 \approx -0.56$ with the transition redshift $z_t \approx 0.66$ \cite{planck_collaboration/2020}. Therefore, the behavior of our $f(Q)$ model indicates a deviation from that of $\Lambda$CDM. Further, the present value of EoS parameter i.e. $w_0=-0.75^{+0.0007}_{-0.0008}$ for the dark energy fluid part indicating a quintessence type behavior of our model. Thus the acquired results indicate that our cosmological $f(Q)$ model well establishes the requirements to describe late-time cosmic acceleration without invoking any dark energy part in the matter content. The cosmological $f(Q)$ model considered in this article have great significance. A power law correction to the STEGR will give rise to branches of solution applicable either to the early universe or to late-time cosmic acceleration \cite{jimenez/2020}. Our cosmological $f(Q)$ model which is nothing but STEGR with constant term, provides a correction to the late-time cosmology, where they can give rise to the dark energy. Also our $f(Q)$ model agrees with the investigation done in \cite{WK}, which indicated that the universe evolves from a radiation dominated epoch towards a matter dominated epoch and eventually settles to an accelerated dark energy dominated epoch.

\section*{Acknowledgments} \label{sec10}
RS acknowledges University Grants Commission(UGC), New Delhi, India for awarding Junior Research Fellowship (UGC-Ref. No.: 191620096030). AD is supported by the grant FRGS/1/2021/STG06/UTAR/02/1. SM acknowledges Department of Science and Technology (DST), Govt. of India, New Delhi, for awarding Senior Research Fellowship (File No. DST/INSPIRE Fellowship/2018/IF18D676).  PKS acknowledges CSIR, New  Delhi, India for financial support to carry out the Research project [No.03(1454)/19/EMR-II, Dt. 02/08/2019]. We are very much grateful to the honorable referees and to the editor for the illuminating suggestions that have significantly improved our work in terms of research quality, and presentation.


\end{document}